\documentclass[twocolumn,superscriptaddress,preprintnumbers,amsmath,amssymb,prl]{revtex4}
\usepackage{amsmath}
\usepackage{amssymb}
\usepackage{dcolumn}
\usepackage{graphicx}
\usepackage{bm}
\usepackage{epstopdf}
\usepackage{threeparttable}
\usepackage{float}
\usepackage{natbib}

\usepackage{appendix}

\begin{document}

\title{The glass-forming ability of model metal-metalloid alloys}
\author{Kai Zhang}
\affiliation{Department of Mechanical Engineering and Materials Science, Yale University, New Haven, Connecticut, 06520, USA}
\affiliation{Center for Research on Interface Structures and Phenomena, Yale University, New Haven, Connecticut, 06520, USA}
\author{Yanhui Liu}
\affiliation{Department of Mechanical Engineering and Materials Science, Yale University, New Haven, Connecticut, 06520, USA}
\affiliation{Center for Research on Interface Structures and Phenomena, Yale University, New Haven, Connecticut, 06520, USA}

\author{Jan Schroers}
\affiliation{Department of Mechanical Engineering and Materials Science, Yale University, New Haven, Connecticut, 06520, USA}
\affiliation{Center for Research on Interface Structures and Phenomena, Yale University, New Haven, Connecticut, 06520, USA}
\author{Mark D. Shattuck}
\affiliation{Department of Physics and Benjamin Levich Institute, The City College of the City University of New York, New York, New York, 10031, USA}
\affiliation{Department of Mechanical Engineering and Materials Science, Yale University, New Haven, Connecticut, 06520, USA}
\author{Corey S. O'Hern}
\affiliation{Department of Mechanical Engineering and Materials Science, Yale University, New Haven, Connecticut, 06520, USA}
\affiliation{Center for Research on Interface Structures and Phenomena, Yale University, New Haven, Connecticut, 06520, USA}
\affiliation{Department of Physics, Yale University, New Haven, Connecticut, 06520, USA}
\affiliation{Department of Applied Physics, Yale University, New Haven, Connecticut, 06520, USA}

\date{\today}

\begin{abstract}

Bulk metallic glasses (BMGs) are amorphous alloys with desirable
mechanical properties and processing capabilities. To date, the design
of new BMGs has largely employed empirical rules and trial-and-error
experimental approaches. {\it Ab initio} computational methods are
currently prohibitively slow to be practically used in searching the
vast space of possible atomic combinations for bulk glass formers.
Here, we perform molecular dynamics simulations of a coarse-grained,
{\it anisotropic} potential, which mimics interatomic covalent bonding, to
measure the critical cooling rates for metal-metalloid alloys as a
function of the atomic size ratio $\sigma_S/\sigma_L$ and number
fraction $x_S$ of the metalloid species.  We show that the regime in
the space of $\sigma_S/\sigma_L$ and $x_S$ where well-mixed, optimal
glass formers occur for patchy and LJ particle mixtures coincides with
that for experimentally observed metal-metalloid glass formers. Our 
simple computational model provides the capability to perform 
combinatorial searches to identify novel glass-forming alloys.

\end{abstract}

\maketitle

\section*{Introduction}
Bulk metallic glasses
(BMGs) are metallic alloys that form amorphous phases with
advantageous material properties~\cite{chen} such as enhanced  strength and elasticity
compared to conventional alloys~\cite{greer:2007} and thermal plastic processing
capabilities that rival those used for
polymers~\cite{schroers}. Despite enormous progress over the past $30$
years in the development and fabrication of BMGs, their commercial use
has been limited due the high cost of some of the constituent elements
and thickness constraints imposed by required rapid cooling.  The search space
for potential new BMGs is vast with roughly $46$ transition metal,
metalloid, and non-metal elements, which give rise to roughly $10^3$,
$10^4$, and $10^5$ candidate binary, ternary, and quaternary alloys,
respectively.

Bulk metallic glass formers can be divided into two primary classes:
metal-metal ({\it i.e.}  transition metal-transition metal) and
metal-metalloid ({\it i.e.} transition metal-metalloid) systems. The
structural and mechanical
properties~\cite{miracle_2004,sheng:2006,cheng_2008} and glass-forming
ability (GFA)~\cite{inoue:2000} of metal-metal systems are much better
understood than for metal-metalloid systems. Dense atomic packing is
the key physical mechanism that determines the glass forming ability
in metal-metal
systems~\cite{miracle_2004,sheng:2006,cheng_2008,busch,schroers}, and
thus these systems have been accurately modeled using coarse-grained,
isotropic hard-sphere and Lennard-Jones interaction
potentials~\cite{zhang:2013,zhang:2014}.  In contrast, since metalloid
atoms form pronounced covalent interatomic bonds~\cite{baskes}, the
atomic structure that influences glass formation is not simply
described by packing efficiency of spherical atoms~\cite{guan:2012}.
Faithfully describing covalent bonding in simulations is
challenging. {\it Ab-initio} simulations can describe covalent bonding
accurately~\cite{ryu:2010}, but {\it ab-initio} simulations beyond
tens of atoms in amorphous structures are not currently
possible. Another possibility is simulations of embedded atom
models that include pairwise interactions and energetic
contributions from electron charge densities~\cite{eam,baskes}.  We take
a simpler geometric computational approach, where we model the covalent
characteristics of metalloid atoms by arranging attractive patches on
the surface of spherical particles to consider the directionality in
covalently bonded structures.  This {\it patchy particle model} has also
been employed to study liquid stability~\cite{smallenburg:2013},
formation of quasicrystals~\cite{doye:2007}, protein
crystallization~\cite{fusco:2013}, and colloidal
self-assembly~\cite{mao:2013,wang:2012}.

Here, we perform molecular dynamics (MD) simulations of the patchy particle
model with $z=4,6,8$, and $12$ patches per particle that yield
diamond, simple cubic, body-centered cubic (BCC), and face-centered
cubic (FCC) lattices in the crystalline state. We thermally quench
equilibrated liquids to low temperature at various cooling rates and
measure the critical cooling rate $R_c$ for each system. We show that
the maximum GFA (minimal $R_c$) for patchy and LJ particle mixtures as
a function of the atomic size ratio $\sigma_S/\sigma_L$ and number
fraction of the metalloid component $x_S$ coincides with the region
where metal-metalloid glass-formers are observed in
experiments~\cite{Ballone,Takeuchi}.  We also used the patchy particle
model to investigate the GFA in systems that form intermetallic
compounds~\cite{abe:2007} since they typically do not possess FCC
symmetry.

\section*{Results}
In previous
work~\cite{zhang:2014}, we showed that the slowest critical cooling
rates for binary hard sphere systems occur in the range $0.8 \gtrsim \sigma_S/\sigma_L \gtrsim 0.73$ and $0.8 \gtrsim x_S \gtrsim 0.5$,
which coincides with the parameters for experimentally observed
metal-metal binary BMGs, such as NiNb, CuZr, CuHf, and CaAl~\cite{xu:2004,xia:2006,wang:2010}.  Similar
results hold for dense binary Lennard-Jones glasses with isotropic
interatomic potentials~\cite{zhang:2013}.  In contrast, the
metal-metalloid glass formers AuSi, PdSi, PtSi, and FeB occur at
smaller $\sigma_S/\sigma_L$ and $x_S$~\cite{lu:2002}. We present results  
from MD simulations that quanitfy the 
glass-forming ability of patchy particles as a function of the number
of patches, their size, and placement on the sphere surface to model
the GFA of metal-metalloid binary glass formers.  (See the Methods 
section.)

We first consider monodisperse systems with $z=12$ patches per
particle and FCC symmetry and measure the average bond orientational
order parameter $\langle Q_6 \rangle$ versus cooling rate $R$ (using 
protocol $1$ in the Methods section) for
several patch sizes $\delta$.  For each $\delta$, $\langle Q_6\rangle
(R)$ is sigmoidal with a midpoint that defines the critical cooling
rate $R_c$. As $R$ decreases toward $R_c$, systems with $z=12$ form
ordered Barlow packings~\cite{barlow:1883} and $\langle Q_6\rangle$
begins to increase as shown in Figure 1. 
In the $\delta
\rightarrow 0$ limit, $R_c$ converges to that for the
Weeks-Chandler-Andersen (WCA) purely repulsive
potential~\cite{weeks:1971}.  As the patch size increases, the $12$
attractive patches promote the formation of FCC nuclei and $R_c$
increases.  For $\delta \gtrsim 0.05$ when patches begin to overlap,
$R_c$ begins to decrease because nucleation and growth of FCC clusters
is frustrated by the concomitant formation of BCC and other types of nuclei.  For
sufficiently large $\delta$, $R_c$ converges to that for Lennard-Jones
(LJ) systems.  This nonmonotonic behavior for $R_c$ versus $\delta$
occurs for other $z$ as well. 

\begin{figure}
\includegraphics[width=3.4in]{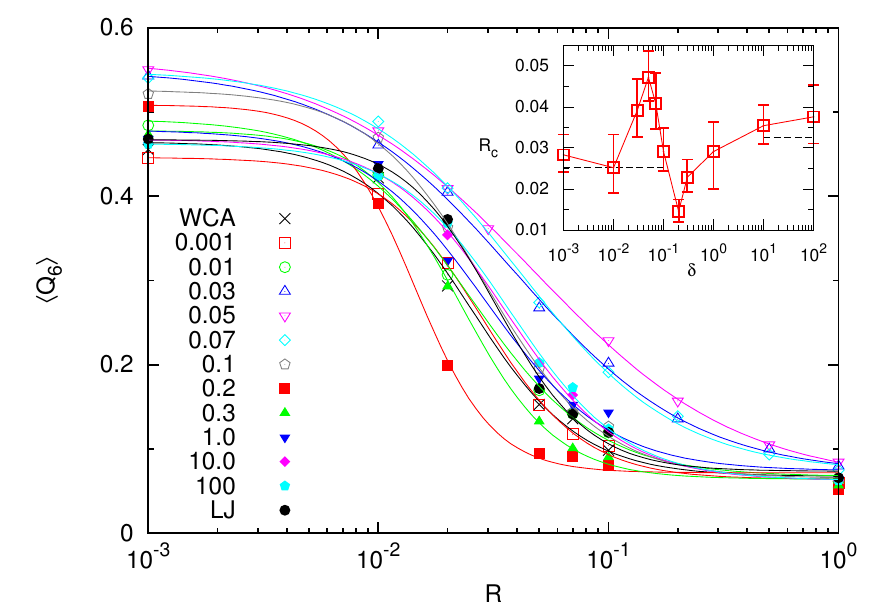}
\caption{{\bf The dependence of GFA on patch size
$\delta$}. The bond orientational order parameter $\langle Q_6 \rangle$
versus cooling rate $R$ for monodisperse patchy particles with $z=12$
cooled at fixed reduced density $\rho^*=1$ for several patch sizes
$\delta$.  $\langle Q_6 \rangle$ was averaged over $96$ separate
trajectories with different initial conditions.  For each $\delta$,
$\langle Q_6 \rangle (R)$ was fit to a logistic function, whose midpoint
gives the critical cooling rate $R_c$. The inset shows $R_c$ versus
$\delta$. The dashed horizontal lines give $R_c$ as the patchy
particle potential approaches either the LJ ($\delta \rightarrow
\infty$) or WCA ($\delta \rightarrow 0$) limiting forms.}
\label{fig:Q6Z12}
\end{figure}

We now investigate the glass-forming ability at fixed patch size
$\delta = 0.1$ as a function of the number and placement of the
patches for $z=4$, $6$, $8$, and $12$, which allows us to tune the
crystalline phase that competes with glass formation.  The GFA for
$z=12$ and $8$ is similar.  As shown in Figure 2a, 
 $\langle Q_6 \rangle$ begins to increase for $R <
R_c \approx 0.04$ with the formation of FCC and BCC clusters for
$z=12$ and $8$, respectively. $\langle Q_4\rangle$ displays a much
more modest change over the same range of $R$. For $z=4$, the glass
competes with the formation of two interpenetrating diamond
lattices~\cite{proserpio} (Figure 2b,c), 
which can be detected using either $\langle Q_6 \rangle$ or $\langle
Q_4\rangle$.  For $z=6$, the simple cubic (SC) phase first forms as
$R$ decreases (indicated by a strong increase in $\langle
Q_4\rangle$), but as $R$ continues to decrease BCC coexists with SC
order (Figure 2c), 
which causes $\langle Q_4 \rangle$ to decrease and $\langle Q_6 \rangle$ to
increase.  In addition, we find that systems for which the competing
crystals are more open possess lower $R_c$.
 
To model metal-metalloid glass formers, we study binary mixtures of
isotropic LJ particles (large metal species) and $z = 4$ patchy
particles (small metalloid species). We chose patchy particles with
tetragonal symmetry to represent silicon atoms since they often
interact with other atoms with four valence electrons in $sp^3$
hybridization orbitals.  In Figure 3a, 
we show a contour plot of the critical cooling rate $R_c$ (obtained by
measuring $\langle Q_6 \rangle (R)$) as a function of
$\sigma_S/\sigma_L$ and $x_S$.  We find two regions along the lines
$x_S \sim 0.2$ and $0.8$ with small values for $R_c$ as determined by
global measures of $\langle Q_6 \rangle$.  However, it is also important to
determine whether the patchy and LJ particles are uniformly mixed at the 
patchy particle number fractions $x_S
\sim 0.2$ and $0.8$.

In Figure 3b, we characterize the
solubility of the patchy particles within the matrix of LJ particles
in glassy states created by rapid cooling to $T_f$ using protocol $2$
in the Methods section.  To quantify the solubility, for each
configuration, we first determine the largest connected cluster of
$N_c$ patchy particles that share faces of Voronoi polyhedra. We then
calculate the radius $R_c$ of the sphere that $N_c$ patchy particles
would assume when confined to a sphere of volume $4 \pi R_c^3/3=
N_c/\rho_S$ at density $\rho_S = N_S / V_S$, where $V_S = V
x_S\sigma_S^3 /(x_L\sigma_L^3+x_S\sigma_S^3)$ and $V$ is the volume of
the cubic simulation cell. $N_{sc}$ is the maximum number of patchy
particles that can be enclosed by a sphere of radius $R_c$ among all
possible locations centered at each of the $N_c$ patchy particles and
the patchy particle solubility $f_S = N_{sc}/N_S$ can be defined for each configuration.  Small
values of $f_S$ indicate that patchy particles are more likely to be
neighbors with LJ particles, not other patchy particles, while $f_S\sim 1$
indicates all patchy particles are in a spherical aggregate (Supplementary information).

Although the global bond orientational order parameter $\langle Q_6
\rangle$ indicates good glass-forming ability for LJ and patchy
particle mixtures at both small ($x_S\sim 0.2$) and large ( $x_S \sim
0.8$) fraction of patchy particles, we find that strong demixing of
the patchy and LJ particles occurs for $x_S \sim 0.8$.  Thus, when
taken together, Figure 3a,b 
show that there is only one region in the $\sigma_S/\sigma_L$ and $x_S$ plane
where well-mixed, good glass-formers occur: $0.2 \lesssim x_S \lesssim
0.4$ and $0.5 \lesssim \sigma_S/\sigma_L \lesssim  0.75$. This region in the
$\sigma_S/\sigma_L$ and $x_S$ plane coincides with the region where
binary metal-metalloid glass alloys ({\it e.g.} AuSi, PdSi, PtSi, and
FeB) are observed. We also find similar simulation results for
mixtures of tri-valent ($z=3$) patchy and LJ particles, which mimic
FeB glass-formers.  In addition, the fact that ternary
metal-metal-metalloid glass formers (CoMnB, FeNiB, FeZrB, and NiPdP),
for which the metal components have similar atomic sizes, also possess
metalloid number fractions $x_S \sim 0.2$ supports our results~\cite{Takeuchi}.

\begin{figure}
\includegraphics[width=3.5in]{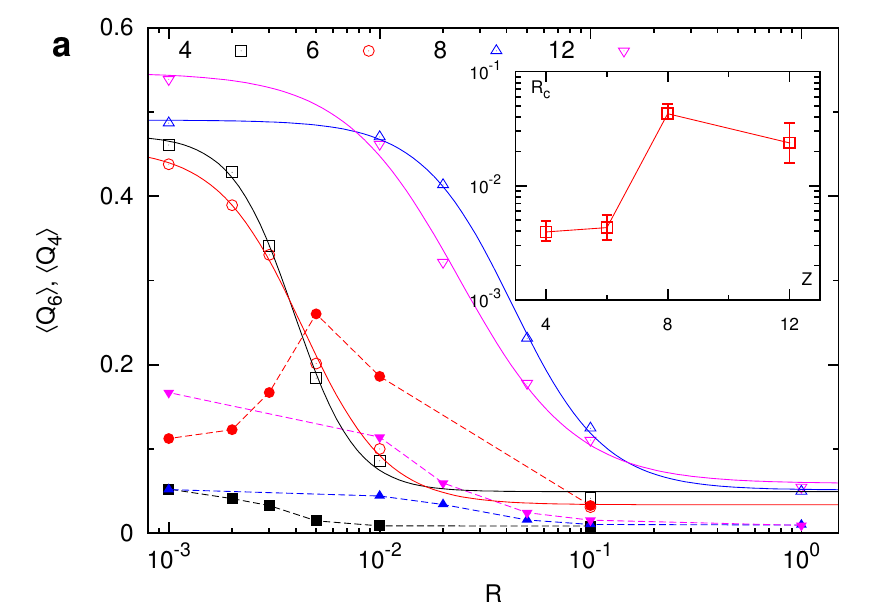}
\includegraphics[width=1.65in]{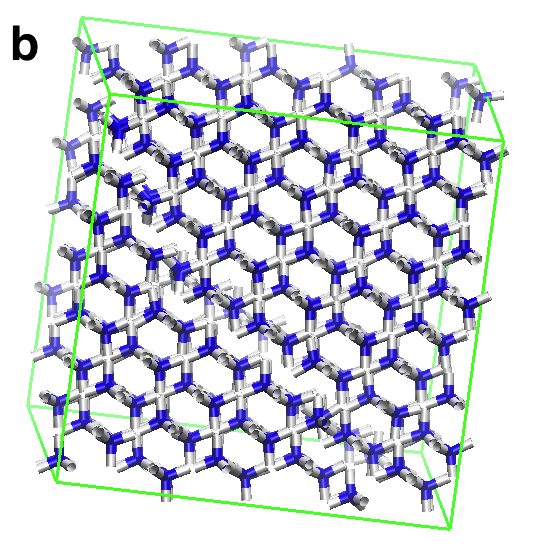}
\includegraphics[width=1.65in]{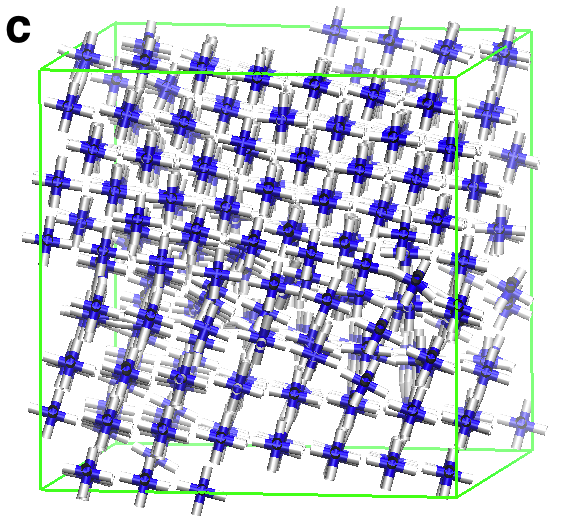}
\caption{ {\bf The dependence of GFA on patch number
$z$}. {\bf a,} Average bond orientational order parameters $\langle Q_6
\rangle$ (open symbols) and $\langle Q_4 \rangle$ (filled symbols)
versus cooling rate $R$ for monodisperse patchy particles with $z=4$
(squares), $6$ (circles), $8$ (upward triangles), and $12$ (downward
triangles) and patch size $\delta=0.1$.  {\bf b,c,} Ordered configurations of
patchy particles in bond representation with particles colored blue and
patches white: interpenetrating diamond lattices for $z=4$ ({\bf b}) and
coexistence of simple cubic and BCC lattices for $z=6$
({\bf c}).}
\label{fig:Q6Zn}
\end{figure}

\begin{figure}
\includegraphics[width=3.5in]{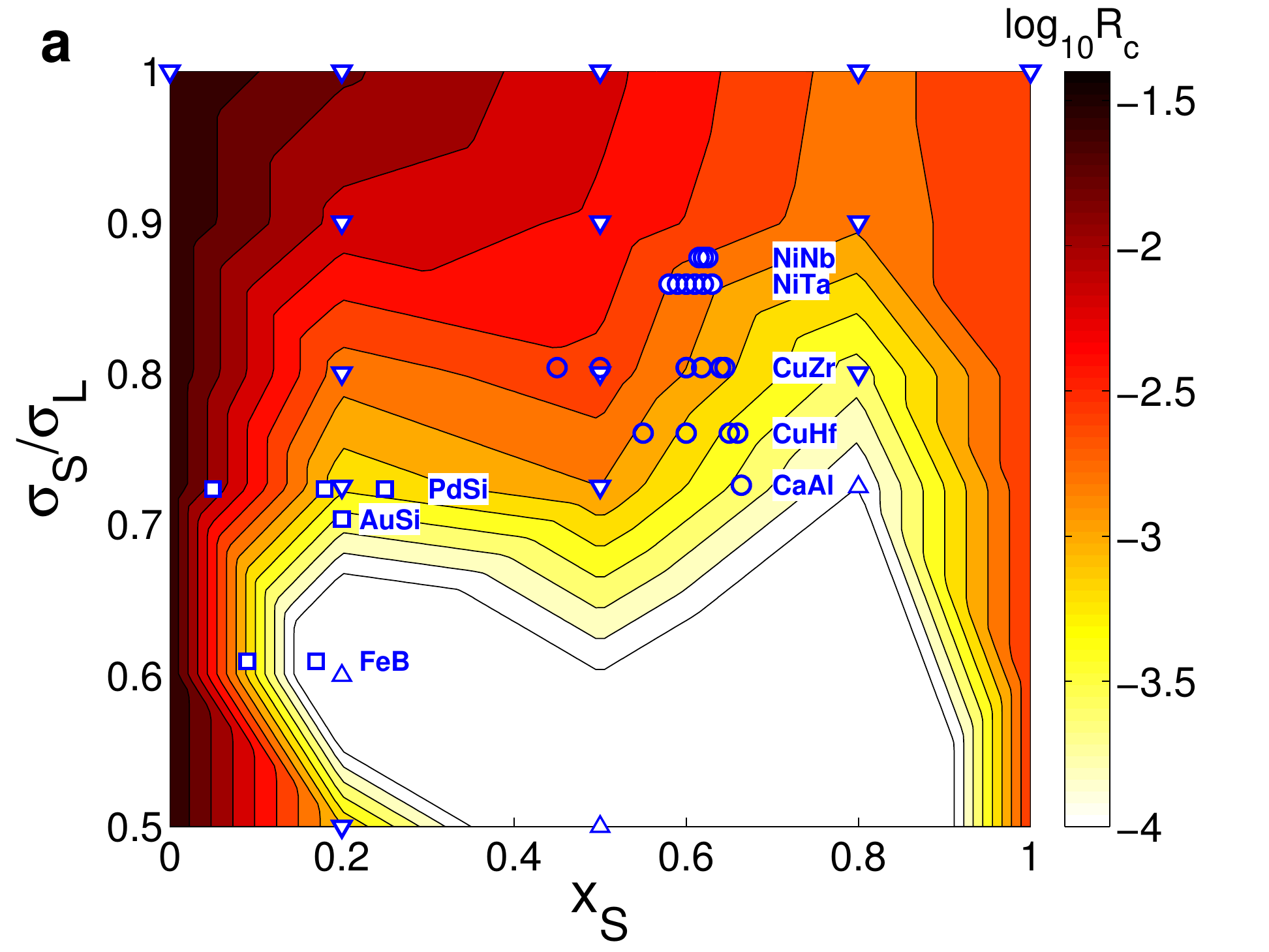}
\includegraphics[width=3.5in]{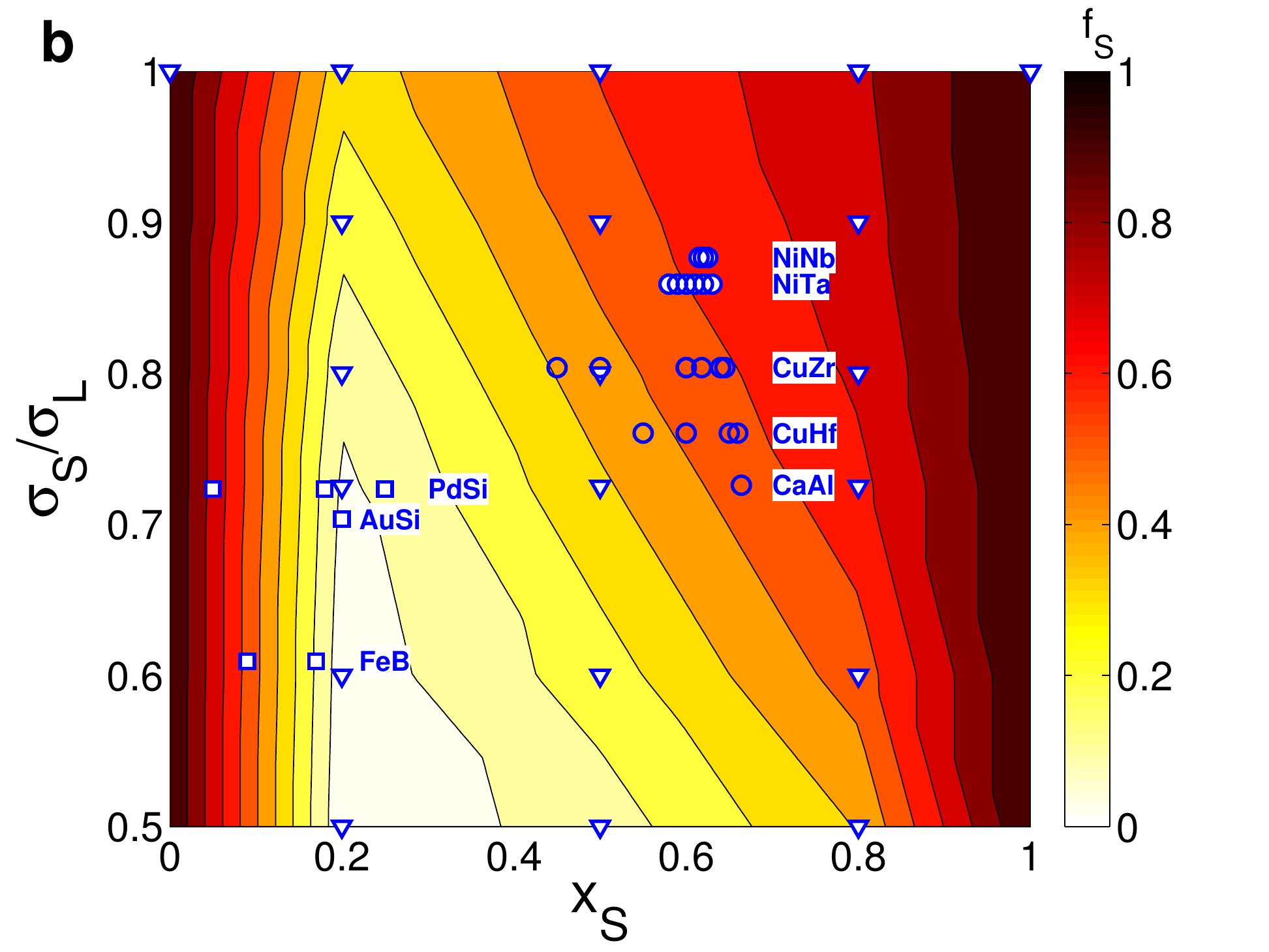}
\caption{{\bf GFA in metal-metalloid systems}. {\bf a,} Contour plot of the critical cooling rate $R_c$ versus
size ratio $\sigma_S/\sigma_L$ and small particle number fraction $x_S$ for a
binary system composed of isotropic (large) LJ particles and (small)
patchy particles with $z=4$ and $\delta = 0.1$. Contours are interpolated using roughly
$20$ MD simulations (downward triangles) spread over parameter
space. Known metal-metal and metal-metalloid binary glass-formers are
indicated by circles and squares, respectively.  {\bf b,}  Measure of the
solubility ($f_S$) of patchy particles within the patchy and LJ particle
mixtures. Number fraction $f_S$ of patchy particles that occur in the largest
connected cluster of patchy particles from glassy configurations 
generated at fast cooling rates ($R=0.1$).}
\label{fig:ZA0ZB4}
\end{figure}

It is also difficult to capture the formation of intermetallic
compounds that possess particular atomic stoichiometries in each local
environment using isotropic hard-sphere or Lennard-Jones potentials.
We show that crystallization of intermetallic compounds can be studied
efficiently using binary mixtures of patchy particles.  We focus on
two model intermetallic compounds: (1) an $AB$ compound with BCC
symmetry and (2) an $AB_2$ compound composed of hexagonal layers.  We
model the $AB$ compound using a binary mixture of $z_S=z_L = 8$ patchy
particles with diameter ratio $\sigma_S/\sigma_L=0.8$ (Figure 4b).  For the $AB_2$
compound, we consider a binary mixture of $z_L=12$ and $z_S = 6$
patchy particles with $\sigma_S/\sigma_L=0.5$ (Figure 4c).  To encourage compound
formation, we only include attractive interactions between patches on
different particle species (with $\delta = 0.1$) and repulsive LJ
interactions between particles of the same type. We find that the
critical cooling rate $R_c$ has a local maximum (and glass-forming
ability has a minimum) at the number fraction expected for compound
formation ($x_S=0.5$ for $AB$ and $x_S=2/3$ for $AB_2$) (Figure 4a). For both $AB$
and $AB_2$, $x_S\sim 0.2$ has the smallest critical cooling rate.
These results emphasize that searches for good glass-formers should avoid 
$x_S$ and $\sigma_S/\sigma_L$ combinations that yield intermetallic compound formation, which can be both stable or metastable.

\begin{figure}
\includegraphics[width=3.5in]{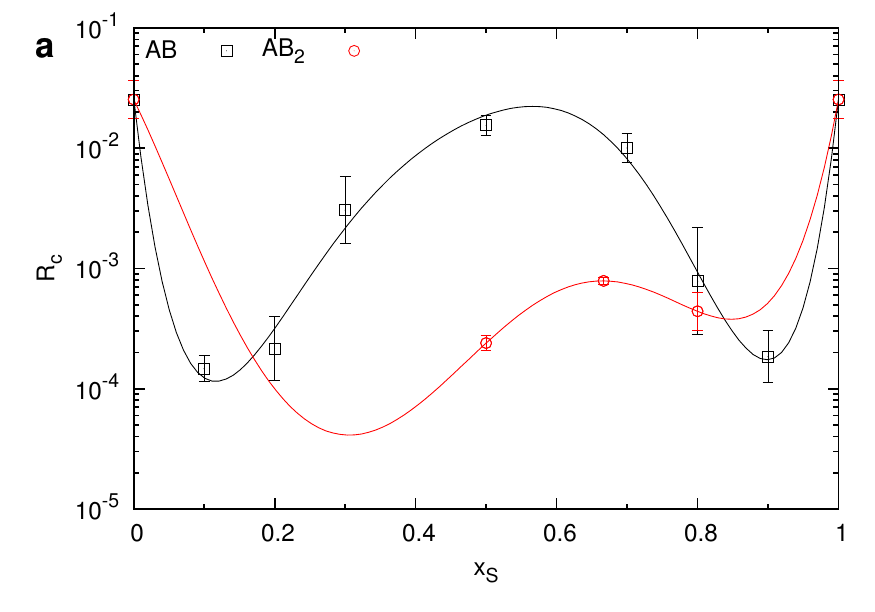}
\includegraphics[width=1.65in]{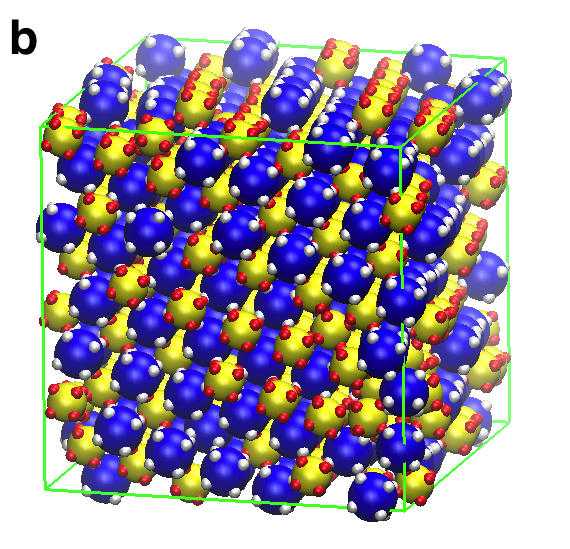}
\includegraphics[width=1.6in]{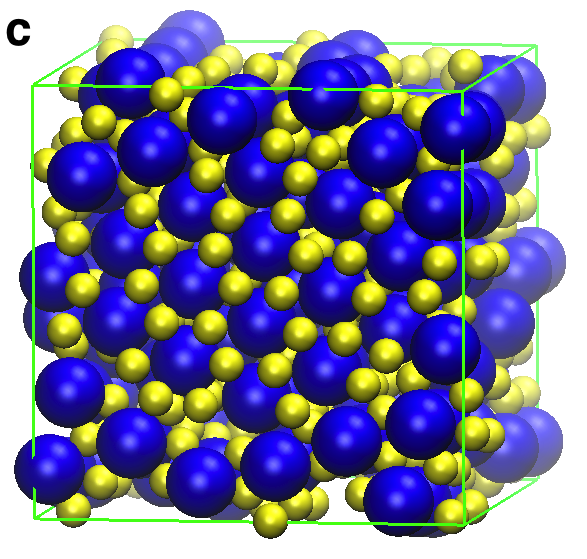}
\caption{ {\bf GFA in compound-forming systems}. {\bf a,} Critical cooling rate $R_c$ versus $x_S$ for model $AB$
(squares) and $AB_2$ (circles) intermetallic compounds. {\bf b,c,}
Intermetallic compounds formed at cooling rate (protocol $2$)
$R=10^{-3} <R_c$.  {\bf b,} $AB$ compound with $z_L=z_S=8$ (patches are
shown as small white and red bumps), BCC symmetry, and
$\sigma_S/\sigma_L=0.8$. {\bf c,} $AB_2$ compound with $z_L=12$ and
$z_S=6$ (patches not shown), stacked hexagonal planes, and
$\sigma_S/\sigma_L = 0.5$. }
\label{fig:compound}
\end{figure}

\section*{Summary}
We performed molecular
dynamics simulations to measure the critical cooling rate $R_c$ and
assess the glass-forming ability (GFA) of patchy and LJ particle
mixtures. We found several key results.  First, we identified
nonmonotonic behavior in $R_c$ as a function of the patch size
$\delta$, indicating a competition between sphere reorientation 
and dense sphere packing in determining GFA in the patchy particle 
model. Second, we tuned the number of patches per particle $z$ and their
placement on the sphere surface to vary the symmetry of the
crystalline phase that competes with glass formation. We found that
systems with more open lattice structures possess lower critical
cooling rates. Third, we showed that the region of $\sigma_S/\sigma_L$
and $x_S$ parameter space where well-mixed, optimal glass-forming LJ
and patchy particle mixtures occur coincides with the region where
metal-metalloid glass-formers are experimentally observed.  In
particular, the number fraction of the metalloid species is small $x_S
\sim 0.2$. The patchy particle model can also be employed to mimic the
formation of intermetallic compounds, and our results emphasize that
searches for good glass-formers should focus on stoichiometries that
do not favor compound formation.  

The search for new BMGs has largely been performed using empirical
rules~\cite{inoue:1998,humerothery:1950} and trial-and-error
experimental techniques~\cite{ding:2014}. Thus, only a small fraction
of the search space of atomic species has been explored with fewer
than $100$ observed BMGs to date~\cite{greer:2009}. Our simple
computational model for metal and metalloid atomic species provides
the capability to perform more efficient and exhaustive combinatorial
searches to identify novel ternary, quaternary, and multi-component
glass-forming alloys.  The smaller set of alloys that are predicted
from simulations to possess slow critical cooling rates can then be
tested experimentally using combinatorial sputtering~\cite{deng:2007}
and other high-throughput BMG characterization
techniques~\cite{li:2008}.

\section*{Methods}
We performed molecular dynamics
simulations~\cite{evans} in a cubic box with volume $V$ of $N$
spherical particles of mass $m$ decorated with $z$ circular disks or
`patches' arranged on the sphere surface with a particular symmetry.
Aligned patches experience Lennard-Jones (LJ) attractive interactions,
whereas the particles interact via short-range repulsions when
patches are not aligned.  The patchy particles are bidisperse with
diameter ratio $\sigma_S/\sigma_L < 1$ and number fraction of
small particles $x_S$.

The interaction potential between patchy particles $i$ and $j$ includes 
an isotropic short-range repulsive interaction and an anisotropic 
attractive interaction between patches~\cite{kern:2003}:
\begin{equation}
\label{split}
u(r_{ij},{\vec s}_{i\alpha},{\vec s}_{j\beta}) = u_R(r_{ij}) +u_A(r_{ij}) 
v(\psi_{i\alpha},\psi_{j\beta}), 
\end{equation}
where $r_{ij}$ is the separation between particles $i$ and $j$,
$u_R(r_{ij})$ is the Weeks-Chandler-Andersen (WCA) purely repulsive
potential~\cite{weeks:1971}, $u_A(r_{ij})$ is the attractive part of
the Lennard-Jones potential truncated and shifted so that it is zero
at $r_c = 2.5 \sigma_{ij}$ (Figure 5a), 
the patch
$\alpha$ on particle $i$ has orientation ${\vec s}_{i\alpha} =
(\sigma_i/2) {\hat n}_{i\alpha}$ with surface normal ${\hat
n}_{i\alpha}$, and $\psi_{i\alpha}$ is the angle between ${\vec
r}_{ij}$ and ${\vec s}_{i\alpha}$ (Figure 5b). 
For the patch-patch interaction, we assume
\begin{equation}
\label{aniso}
v(\psi_{i\alpha},\psi_{j\beta}) = e^{-\frac{(1-\cos \psi_{i\alpha})^2}{\delta_{i\alpha}^2}}e^{-\frac{(1-\cos \psi_{j\beta})^2}{\delta_{j\beta}^2}}, 
\end{equation}
which is maximized when $\psi_{i\alpha} = \psi_{j\beta} =
0$. $\delta_{i\alpha}$ gives the width of the interaction for patch
$\alpha$ on particle $i$. For each patch $\alpha$, we only include an
interaction with the patch $\beta$ that has the largest
$v(\psi_{i\alpha},\psi_{j\beta})$.  In the large patch size limit,
equation (\ref{aniso}) becomes isotropic and the patchy particle model
becomes identical to the full Lennard-Jones potential.  In the
opposite limit, as $\delta\rightarrow 0$, the patchy particle
potential reduces to $u_R(r_{ij})$.  We considered particles with
$z=4$, $6$, $8$, and $12$ patches arranged on the sphere surface with
tetrahedral, simple cubic, body-centered cubic (BCC), and
face-centered cubic (FCC) symmetry (inset to Figure 5a). 
For the investigations of $AB_2$ compounds, we also considered
systems with $z_L=12$ and $z_S=6$ for the large and small particles
and arrangements that are compatible with the $AB_2$ symmetry (Supplementary information).

\begin{figure}
\includegraphics[width=3.6in]{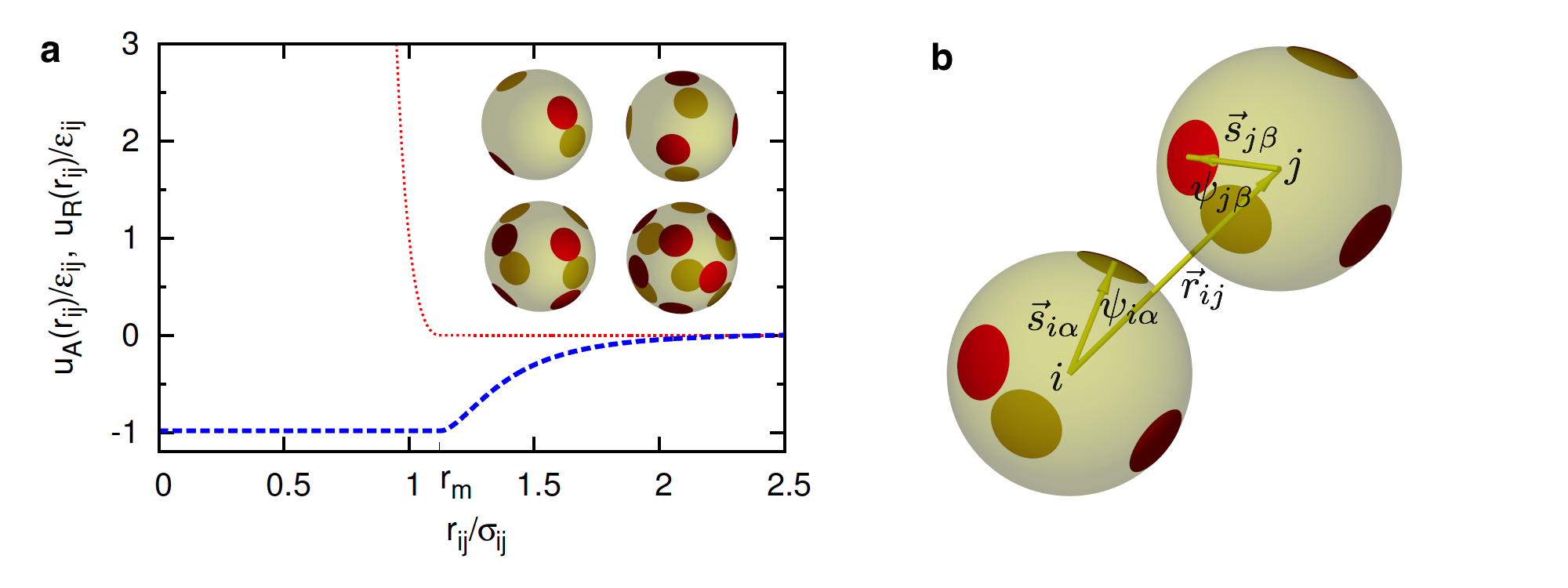}
\caption{ {\bf Definition of the model potential}. {\bf a,} The purely repulsive WCA potential $u_R(r_{ij})$ is zero
for $r_{ij} \ge r_{\rm m} = 2^{1/6}\sigma_{ij}$ and the attractive
part $u_A(r_{ij})$ of the Lennard-Jones potential is truncated and
shifted so that it is zero at $r_c = 2.5 \sigma_{ij}$. Here, the
Lennard-Jones energy parameters are $\epsilon_{SS} /\epsilon_{LL} =
\epsilon_{LS}/\epsilon_{LL} = 1$. The inset shows examples of
particles with $4$, $6$, $8$, and $12$ patches with tetrahedral,
simple cubic, BCC, and FCC symmetry, respectively. Red patches
correspond to those on the front surface of the sphere, while dark
yellow patches indicate those on the back surface. {\bf b,} Definitions 
of quantities in the patchy particle interaction potential in equations (\ref{split})
and (\ref{aniso}).}
\label{fig:potential}
\end{figure}

To assess the glass-forming ability of patchy particle systems, we
measured the critical cooling rate $R_c$ below which crystallization
begins to occur. The systems are cooled using one of two protocols:
(1) the temperature is decreased exponentially in time $T(t) = T_0
e^{- R t}$ at reduced density $\rho^*=N\sigma_L^3/V = 1.0$ from
$T_0/\epsilon_{LL} = 2.0$ in the liquid regime to $T_f /\epsilon_{LL}
= 0.01$ in the glassy state and (2) both the temperature and pressure
$p$ are decreased exponentially in time with $p(t) = p_0 e^{- R_p t}$,
where $R_p = R$, the state point $T_0/\epsilon_{LL}$ and $p_0
\sigma_{LL}^3/\epsilon_{LL}=20$ is in the liquid regime, and the state
point $T_f/\epsilon_{LL}$ and $p_f \sigma_{LL}^3/\epsilon_{LL}=0.1$ is
in the glassy regime. Protocol $2$ was implemented for systems with
$z<12$ to allow the system to choose a box volume most compatible with
the low-energy crystal structure.  The emergence of crystalline order
is signaled by a strong increase of the bond orientational order
parameters $Q_6$ and $Q_4$~\cite{steinhardt:1983} for cooling rates $R
< R_c$.  We focused on systems with $N=500$ particles, but also
studied systems with $N=1372$
to assess finite-size effects (Supplementary information). The dynamics were solved by integrating
Newton's equation of motion for the translation and rotational degrees
of freedom using Gear predictor-corrector methods with time step
$\Delta t = 10^{-3} \sigma_{LL}\sqrt{m/\epsilon_{LL}}
$~\cite{allen:1987}.

\bibliography{patchy}

\begin{thebibliography}{39}
\expandafter\ifx\csname natexlab\endcsname\relax\def\natexlab#1{#1}\fi
\expandafter\ifx\csname bibnamefont\endcsname\relax
  \def\bibnamefont#1{#1}\fi
\expandafter\ifx\csname bibfnamefont\endcsname\relax
  \def\bibfnamefont#1{#1}\fi
\expandafter\ifx\csname citenamefont\endcsname\relax
  \def\citenamefont#1{#1}\fi
\expandafter\ifx\csname url\endcsname\relax
  \def\url#1{\texttt{#1}}\fi
\expandafter\ifx\csname urlprefix\endcsname\relax\def\urlprefix{URL }\fi
\providecommand{\bibinfo}[2]{#2}
\providecommand{\eprint}[2][]{\url{#2}}

\bibitem[{\citenamefont{Chen}(1980)}]{chen}
\bibinfo{author}{\bibfnamefont{H.~S.} \bibnamefont{Chen}},
  \bibinfo{journal}{Rep. Prog. Phys.} \textbf{\bibinfo{volume}{43}},
  \bibinfo{pages}{353} (\bibinfo{year}{1980}).

\bibitem[{\citenamefont{Greer and Ma}(2007)}]{greer:2007}
\bibinfo{author}{\bibfnamefont{A.~L.} \bibnamefont{Greer}} \bibnamefont{and}
  \bibinfo{author}{\bibfnamefont{E.}~\bibnamefont{Ma}}, \bibinfo{journal}{MRS
  Bulletin} \textbf{\bibinfo{volume}{32}}, \bibinfo{pages}{611}
  (\bibinfo{year}{2007}).

\bibitem[{\citenamefont{Schroers}(2013)}]{schroers}
\bibinfo{author}{\bibfnamefont{J.}~\bibnamefont{Schroers}},
  \bibinfo{journal}{Physics Today} \textbf{\bibinfo{volume}{66}},
  \bibinfo{pages}{32} (\bibinfo{year}{2013}).

\bibitem[{\citenamefont{Miracle}(2004)}]{miracle_2004}
\bibinfo{author}{\bibfnamefont{D.~B.} \bibnamefont{Miracle}},
  \bibinfo{journal}{Nature Materials} \textbf{\bibinfo{volume}{3}},
  \bibinfo{pages}{697} (\bibinfo{year}{2004}).

\bibitem[{\citenamefont{Sheng et~al.}(2006)\citenamefont{Sheng, Luo, Alamgir,
  Bai, and Ma}}]{sheng:2006}
\bibinfo{author}{\bibfnamefont{H.~W.} \bibnamefont{Sheng}},
  \bibinfo{author}{\bibfnamefont{W.~K.} \bibnamefont{Luo}},
  \bibinfo{author}{\bibfnamefont{F.~M.} \bibnamefont{Alamgir}},
  \bibinfo{author}{\bibfnamefont{J.~M.} \bibnamefont{Bai}}, \bibnamefont{and}
  \bibinfo{author}{\bibfnamefont{E.}~\bibnamefont{Ma}},
  \bibinfo{journal}{Nature} \textbf{\bibinfo{volume}{439}},
  \bibinfo{pages}{419} (\bibinfo{year}{2006}).

\bibitem[{\citenamefont{Cheng et~al.}(2008)\citenamefont{Cheng, Sheng, and
  Ma}}]{cheng_2008}
\bibinfo{author}{\bibfnamefont{Y.~Q.} \bibnamefont{Cheng}},
  \bibinfo{author}{\bibfnamefont{H.~W.} \bibnamefont{Sheng}}, \bibnamefont{and}
  \bibinfo{author}{\bibfnamefont{E.}~\bibnamefont{Ma}}, \bibinfo{journal}{Phys.
  Rev. B} \textbf{\bibinfo{volume}{78}}, \bibinfo{pages}{014207}
  (\bibinfo{year}{2008}).

\bibitem[{\citenamefont{Inoue}(2000)}]{inoue:2000}
\bibinfo{author}{\bibfnamefont{A.}~\bibnamefont{Inoue}}, \bibinfo{journal}{Acta
  Mater.} \textbf{\bibinfo{volume}{48}}, \bibinfo{pages}{279}
  (\bibinfo{year}{2000}).

\bibitem[{\citenamefont{Busch et~al.}(2007)\citenamefont{Busch, Schroers, and
  Wang}}]{busch}
\bibinfo{author}{\bibfnamefont{R.}~\bibnamefont{Busch}},
  \bibinfo{author}{\bibfnamefont{J.}~\bibnamefont{Schroers}}, \bibnamefont{and}
  \bibinfo{author}{\bibfnamefont{W.~H.} \bibnamefont{Wang}},
  \bibinfo{journal}{MRS Bulletin} \textbf{\bibinfo{volume}{32}},
  \bibinfo{pages}{620} (\bibinfo{year}{2007}).

\bibitem[{\citenamefont{Zhang et~al.}(2013)\citenamefont{Zhang, Wang,
  Papanikolaou, Liu, Schroers, Shattuck, and O'Hern}}]{zhang:2013}
\bibinfo{author}{\bibfnamefont{K.}~\bibnamefont{Zhang}},
  \bibinfo{author}{\bibfnamefont{M.}~\bibnamefont{Wang}},
  \bibinfo{author}{\bibfnamefont{S.}~\bibnamefont{Papanikolaou}},
  \bibinfo{author}{\bibfnamefont{Y.}~\bibnamefont{Liu}},
  \bibinfo{author}{\bibfnamefont{J.}~\bibnamefont{Schroers}},
  \bibinfo{author}{\bibfnamefont{M.~D.} \bibnamefont{Shattuck}},
  \bibnamefont{and} \bibinfo{author}{\bibfnamefont{C.~S.}
  \bibnamefont{O'Hern}}, \bibinfo{journal}{J. Chem. Phys.}
  \textbf{\bibinfo{volume}{139}}, \bibinfo{pages}{124503}
  (\bibinfo{year}{2013}).

\bibitem[{\citenamefont{Zhang et~al.}(2014)\citenamefont{Zhang, Smith, Wang,
  Liu, Schroers, Shattuck, and O'Hern}}]{zhang:2014}
\bibinfo{author}{\bibfnamefont{K.}~\bibnamefont{Zhang}},
  \bibinfo{author}{\bibfnamefont{W.~W.} \bibnamefont{Smith}},
  \bibinfo{author}{\bibfnamefont{M.}~\bibnamefont{Wang}},
  \bibinfo{author}{\bibfnamefont{Y.}~\bibnamefont{Liu}},
  \bibinfo{author}{\bibfnamefont{J.}~\bibnamefont{Schroers}},
  \bibinfo{author}{\bibfnamefont{M.~D.} \bibnamefont{Shattuck}},
  \bibnamefont{and} \bibinfo{author}{\bibfnamefont{C.~S.}
  \bibnamefont{O'Hern}}, \bibinfo{journal}{Phys. Rev. E}
  \textbf{\bibinfo{volume}{90}}, \bibinfo{pages}{032311}
  (\bibinfo{year}{2014}).

\bibitem[{\citenamefont{Baskes}(1987)}]{baskes}
\bibinfo{author}{\bibfnamefont{M.~I.} \bibnamefont{Baskes}},
  \bibinfo{journal}{Phys. Rev. Lett.} \textbf{\bibinfo{volume}{59}},
  \bibinfo{pages}{2666} (\bibinfo{year}{1987}).

\bibitem[{\citenamefont{Guan et~al.}(2012)\citenamefont{Guan, Fujita, Hirata,
  Liu, and Chen}}]{guan:2012}
\bibinfo{author}{\bibfnamefont{P.~F.} \bibnamefont{Guan}},
  \bibinfo{author}{\bibfnamefont{T.}~\bibnamefont{Fujita}},
  \bibinfo{author}{\bibfnamefont{A.}~\bibnamefont{Hirata}},
  \bibinfo{author}{\bibfnamefont{Y.~H.} \bibnamefont{Liu}}, \bibnamefont{and}
  \bibinfo{author}{\bibfnamefont{M.~W.} \bibnamefont{Chen}},
  \bibinfo{journal}{Phys. Rev. Lett.} \textbf{\bibinfo{volume}{108}},
  \bibinfo{pages}{175501} (\bibinfo{year}{2012}).

\bibitem[{\citenamefont{Ryu and Cai}(2010)}]{ryu:2010}
\bibinfo{author}{\bibfnamefont{S.}~\bibnamefont{Ryu}} \bibnamefont{and}
  \bibinfo{author}{\bibfnamefont{W.}~\bibnamefont{Cai}}, \bibinfo{journal}{J.
  Phys.: Condens. Matter} \textbf{\bibinfo{volume}{22}},
  \bibinfo{pages}{055401} (\bibinfo{year}{2010}).

\bibitem[{\citenamefont{Daw and Baskes}(1984)}]{eam}
\bibinfo{author}{\bibfnamefont{M.~S.} \bibnamefont{Daw}} \bibnamefont{and}
  \bibinfo{author}{\bibfnamefont{M.}~\bibnamefont{Baskes}},
  \bibinfo{journal}{Phys. Rev. B} \textbf{\bibinfo{volume}{29}},
  \bibinfo{pages}{6443} (\bibinfo{year}{1984}).

\bibitem[{\citenamefont{Smallenburg and Sciortino}(2013)}]{smallenburg:2013}
\bibinfo{author}{\bibfnamefont{F.}~\bibnamefont{Smallenburg}} \bibnamefont{and}
  \bibinfo{author}{\bibfnamefont{F.}~\bibnamefont{Sciortino}},
  \bibinfo{journal}{Nature Phys.} \textbf{\bibinfo{volume}{9}},
  \bibinfo{pages}{554} (\bibinfo{year}{2013}).

\bibitem[{\citenamefont{Doye et~al.}(2007)\citenamefont{Doye, Louis, Lin, Noya,
  Wilber, Kok, and Lyus}}]{doye:2007}
\bibinfo{author}{\bibfnamefont{J.~P.~K.} \bibnamefont{Doye}},
  \bibinfo{author}{\bibfnamefont{A.~A.} \bibnamefont{Louis}},
  \bibinfo{author}{\bibfnamefont{L.~R.} \bibnamefont{Lin},
  \bibfnamefont{I.~Allen}}, \bibinfo{author}{\bibfnamefont{E.~G.}
  \bibnamefont{Noya}}, \bibinfo{author}{\bibfnamefont{A.~W.}
  \bibnamefont{Wilber}}, \bibinfo{author}{\bibfnamefont{H.~C.}
  \bibnamefont{Kok}}, \bibnamefont{and}
  \bibinfo{author}{\bibfnamefont{R.}~\bibnamefont{Lyus}},
  \bibinfo{journal}{Phys. Chem. Chem. Phys.} \textbf{\bibinfo{volume}{9}},
  \bibinfo{pages}{2197} (\bibinfo{year}{2007}).

\bibitem[{\citenamefont{Fusco and Charbonneau}(2013)}]{fusco:2013}
\bibinfo{author}{\bibfnamefont{D.}~\bibnamefont{Fusco}} \bibnamefont{and}
  \bibinfo{author}{\bibfnamefont{P.}~\bibnamefont{Charbonneau}},
  \bibinfo{journal}{Phys. Rev. E} \textbf{\bibinfo{volume}{88}},
  \bibinfo{pages}{012721} (\bibinfo{year}{2013}).

\bibitem[{\citenamefont{Mao et~al.}(2013)\citenamefont{Mao, Chen, and
  Granick}}]{mao:2013}
\bibinfo{author}{\bibfnamefont{X.}~\bibnamefont{Mao}},
  \bibinfo{author}{\bibfnamefont{Q.}~\bibnamefont{Chen}}, \bibnamefont{and}
  \bibinfo{author}{\bibfnamefont{S.}~\bibnamefont{Granick}},
  \bibinfo{journal}{Nature Mater.} \textbf{\bibinfo{volume}{12}},
  \bibinfo{pages}{217} (\bibinfo{year}{2013}).

\bibitem[{\citenamefont{Wang et~al.}(2012)\citenamefont{Wang, Wang, Breed,
  Manoharan, Feng, Hollingsworth, Weck, and Pine}}]{wang:2012}
\bibinfo{author}{\bibfnamefont{Y.}~\bibnamefont{Wang}},
  \bibinfo{author}{\bibfnamefont{Y.}~\bibnamefont{Wang}},
  \bibinfo{author}{\bibfnamefont{D.}~\bibnamefont{Breed}},
  \bibinfo{author}{\bibfnamefont{V.~N.} \bibnamefont{Manoharan}},
  \bibinfo{author}{\bibfnamefont{L.}~\bibnamefont{Feng}},
  \bibinfo{author}{\bibfnamefont{A.~D.} \bibnamefont{Hollingsworth}},
  \bibinfo{author}{\bibfnamefont{M.}~\bibnamefont{Weck}}, \bibnamefont{and}
  \bibinfo{author}{\bibfnamefont{D.~J.} \bibnamefont{Pine}},
  \bibinfo{journal}{Nature} \textbf{\bibinfo{volume}{491}}, \bibinfo{pages}{51}
  (\bibinfo{year}{2012}).

\bibitem[{\citenamefont{Ballone and Rubini}(1995)}]{Ballone}
\bibinfo{author}{\bibfnamefont{P.}~\bibnamefont{Ballone}} \bibnamefont{and}
  \bibinfo{author}{\bibfnamefont{S.}~\bibnamefont{Rubini}},
  \bibinfo{journal}{Phys. Rev. B} \textbf{\bibinfo{volume}{51}},
  \bibinfo{pages}{14962} (\bibinfo{year}{1995}).

\bibitem[{\citenamefont{Takeuchi and Inoue}(2001)}]{Takeuchi}
\bibinfo{author}{\bibfnamefont{A.}~\bibnamefont{Takeuchi}} \bibnamefont{and}
  \bibinfo{author}{\bibfnamefont{A.}~\bibnamefont{Inoue}},
  \bibinfo{journal}{Mater. Trans.} \textbf{\bibinfo{volume}{42}},
  \bibinfo{pages}{1435} (\bibinfo{year}{2001}).

\bibitem[{\citenamefont{Abe et~al.}(2007)\citenamefont{Abe, Shimono, Ode, and
  Onodera}}]{abe:2007}
\bibinfo{author}{\bibfnamefont{T.}~\bibnamefont{Abe}},
  \bibinfo{author}{\bibfnamefont{M.}~\bibnamefont{Shimono}},
  \bibinfo{author}{\bibfnamefont{M.}~\bibnamefont{Ode}}, \bibnamefont{and}
  \bibinfo{author}{\bibfnamefont{H.}~\bibnamefont{Onodera}},
  \bibinfo{journal}{Journal of Alloys and Compounds}
  \textbf{\bibinfo{volume}{434-435}}, \bibinfo{pages}{152}
  (\bibinfo{year}{2007}).

\bibitem[{\citenamefont{Xu et~al.}(2004)\citenamefont{Xu, Lohwongwatana, Duan,
  Johnson, and Garland}}]{xu:2004}
\bibinfo{author}{\bibfnamefont{D.}~\bibnamefont{Xu}},
  \bibinfo{author}{\bibfnamefont{B.}~\bibnamefont{Lohwongwatana}},
  \bibinfo{author}{\bibfnamefont{G.}~\bibnamefont{Duan}},
  \bibinfo{author}{\bibfnamefont{W.~L.} \bibnamefont{Johnson}},
  \bibnamefont{and} \bibinfo{author}{\bibfnamefont{C.}~\bibnamefont{Garland}},
  \bibinfo{journal}{Acta Materialia} \textbf{\bibinfo{volume}{52}},
  \bibinfo{pages}{2621} (\bibinfo{year}{2004}).

\bibitem[{\citenamefont{Xia et~al.}(2006)\citenamefont{Xia, Li, Fang, Wei, and
  Dong}}]{xia:2006}
\bibinfo{author}{\bibfnamefont{L.}~\bibnamefont{Xia}},
  \bibinfo{author}{\bibfnamefont{W.~H.} \bibnamefont{Li}},
  \bibinfo{author}{\bibfnamefont{S.~S.} \bibnamefont{Fang}},
  \bibinfo{author}{\bibfnamefont{B.~C.} \bibnamefont{Wei}}, \bibnamefont{and}
  \bibinfo{author}{\bibfnamefont{Y.~D.} \bibnamefont{Dong}},
  \bibinfo{journal}{J. Appl. Phys.} \textbf{\bibinfo{volume}{99}},
  \bibinfo{pages}{026103} (\bibinfo{year}{2006}).

\bibitem[{\citenamefont{Wang et~al.}(2010)\citenamefont{Wang, Wang, Zhao, and
  Dong}}]{wang:2010}
\bibinfo{author}{\bibfnamefont{Y.}~\bibnamefont{Wang}},
  \bibinfo{author}{\bibfnamefont{Q.}~\bibnamefont{Wang}},
  \bibinfo{author}{\bibfnamefont{J.}~\bibnamefont{Zhao}}, \bibnamefont{and}
  \bibinfo{author}{\bibfnamefont{C.}~\bibnamefont{Dong}},
  \bibinfo{journal}{Scripta Materialia} \textbf{\bibinfo{volume}{63}},
  \bibinfo{pages}{178} (\bibinfo{year}{2010}).

\bibitem[{\citenamefont{Lu and Liu}(2002)}]{lu:2002}
\bibinfo{author}{\bibfnamefont{Z.~P.} \bibnamefont{Lu}} \bibnamefont{and}
  \bibinfo{author}{\bibfnamefont{C.~T.} \bibnamefont{Liu}},
  \bibinfo{journal}{Acta Materialia} \textbf{\bibinfo{volume}{50}},
  \bibinfo{pages}{3501} (\bibinfo{year}{2002}).

\bibitem[{\citenamefont{Barlow}(1883)}]{barlow:1883}
\bibinfo{author}{\bibfnamefont{W.}~\bibnamefont{Barlow}},
  \bibinfo{journal}{Nature} \textbf{\bibinfo{volume}{29}}, \bibinfo{pages}{186}
  (\bibinfo{year}{1883}).

\bibitem[{\citenamefont{Weeks et~al.}(1971)\citenamefont{Weeks, Chandler, and
  Andersen}}]{weeks:1971}
\bibinfo{author}{\bibfnamefont{J.~D.} \bibnamefont{Weeks}},
  \bibinfo{author}{\bibfnamefont{D.}~\bibnamefont{Chandler}}, \bibnamefont{and}
  \bibinfo{author}{\bibfnamefont{H.~C.} \bibnamefont{Andersen}},
  \bibinfo{journal}{J. Chem. Phys.} \textbf{\bibinfo{volume}{54}},
  \bibinfo{pages}{5237} (\bibinfo{year}{1971}).

\bibitem[{\citenamefont{Proserpio et~al.}(1994)\citenamefont{Proserpio,
  Hoffmann, and Preuss}}]{proserpio}
\bibinfo{author}{\bibfnamefont{D.~M.} \bibnamefont{Proserpio}},
  \bibinfo{author}{\bibfnamefont{R.}~\bibnamefont{Hoffmann}}, \bibnamefont{and}
  \bibinfo{author}{\bibfnamefont{P.}~\bibnamefont{Preuss}},
  \bibinfo{journal}{J. Am. Chem. Soc.} \textbf{\bibinfo{volume}{116}},
  \bibinfo{pages}{9634} (\bibinfo{year}{1994}).

\bibitem[{\citenamefont{Inoue et~al.}(1998)\citenamefont{Inoue, Zhang, and
  Takeuchi}}]{inoue:1998}
\bibinfo{author}{\bibfnamefont{A.}~\bibnamefont{Inoue}},
  \bibinfo{author}{\bibfnamefont{T.}~\bibnamefont{Zhang}}, \bibnamefont{and}
  \bibinfo{author}{\bibfnamefont{A.}~\bibnamefont{Takeuchi}},
  \bibinfo{journal}{Materials Science Forum}
  \textbf{\bibinfo{volume}{269-272}}, \bibinfo{pages}{855}
  (\bibinfo{year}{1998}).

\bibitem[{\citenamefont{Hume-Rothery}(1950)}]{humerothery:1950}
\bibinfo{author}{\bibfnamefont{W.}~\bibnamefont{Hume-Rothery}},
  \emph{\bibinfo{title}{The Structure of Metals and Alloys}}
  (\bibinfo{publisher}{The Institute of Metals, London}, \bibinfo{year}{1950}).

\bibitem[{\citenamefont{Ding et~al.}(2014)\citenamefont{Ding, Liu, Li, Liu,
  Sohn, Walker, and Schroers}}]{ding:2014}
\bibinfo{author}{\bibfnamefont{S.}~\bibnamefont{Ding}},
  \bibinfo{author}{\bibfnamefont{Y.}~\bibnamefont{Liu}},
  \bibinfo{author}{\bibfnamefont{Y.}~\bibnamefont{Li}},
  \bibinfo{author}{\bibfnamefont{Z.}~\bibnamefont{Liu}},
  \bibinfo{author}{\bibfnamefont{S.}~\bibnamefont{Sohn}},
  \bibinfo{author}{\bibfnamefont{F.~J.} \bibnamefont{Walker}},
  \bibnamefont{and} \bibinfo{author}{\bibfnamefont{J.}~\bibnamefont{Schroers}},
  \bibinfo{journal}{Nature Mater.} \textbf{\bibinfo{volume}{13}},
  \bibinfo{pages}{494} (\bibinfo{year}{2014}).

\bibitem[{\citenamefont{Greer}(2009)}]{greer:2009}
\bibinfo{author}{\bibfnamefont{A.~L.} \bibnamefont{Greer}},
  \bibinfo{journal}{Materials Today} \textbf{\bibinfo{volume}{12}},
  \bibinfo{pages}{14 } (\bibinfo{year}{2009}).

\bibitem[{\citenamefont{Deng et~al.}(2007)\citenamefont{Deng, Guan, Fowlkes,
  Wen, Liu, Pharr, Liaw, Liu, and Rack}}]{deng:2007}
\bibinfo{author}{\bibfnamefont{Y.}~\bibnamefont{Deng}},
  \bibinfo{author}{\bibfnamefont{Y.}~\bibnamefont{Guan}},
  \bibinfo{author}{\bibfnamefont{J.}~\bibnamefont{Fowlkes}},
  \bibinfo{author}{\bibfnamefont{S.}~\bibnamefont{Wen}},
  \bibinfo{author}{\bibfnamefont{F.}~\bibnamefont{Liu}},
  \bibinfo{author}{\bibfnamefont{G.}~\bibnamefont{Pharr}},
  \bibinfo{author}{\bibfnamefont{P.}~\bibnamefont{Liaw}},
  \bibinfo{author}{\bibfnamefont{C.}~\bibnamefont{Liu}}, \bibnamefont{and}
  \bibinfo{author}{\bibfnamefont{P.}~\bibnamefont{Rack}},
  \bibinfo{journal}{Intermetallics} \textbf{\bibinfo{volume}{15}},
  \bibinfo{pages}{1208 } (\bibinfo{year}{2007}).

\bibitem[{\citenamefont{Li et~al.}(2008)\citenamefont{Li, Guo, Kalb, and
  Thompson}}]{li:2008}
\bibinfo{author}{\bibfnamefont{Y.}~\bibnamefont{Li}},
  \bibinfo{author}{\bibfnamefont{Q.}~\bibnamefont{Guo}},
  \bibinfo{author}{\bibfnamefont{J.~A.} \bibnamefont{Kalb}}, \bibnamefont{and}
  \bibinfo{author}{\bibfnamefont{C.~V.} \bibnamefont{Thompson}},
  \bibinfo{journal}{Science} \textbf{\bibinfo{volume}{322}},
  \bibinfo{pages}{1816} (\bibinfo{year}{2008}).

\bibitem[{\citenamefont{Evans}(1983)}]{evans}
\bibinfo{author}{\bibfnamefont{D.~J.} \bibnamefont{Evans}},
  \bibinfo{journal}{Chem. Phys.} \textbf{\bibinfo{volume}{77}},
  \bibinfo{pages}{63} (\bibinfo{year}{1983}).

\bibitem[{\citenamefont{Kern and Frenkel}(2003)}]{kern:2003}
\bibinfo{author}{\bibfnamefont{N.}~\bibnamefont{Kern}} \bibnamefont{and}
  \bibinfo{author}{\bibfnamefont{D.}~\bibnamefont{Frenkel}},
  \bibinfo{journal}{J. Chem. Phys.} \textbf{\bibinfo{volume}{118}},
  \bibinfo{pages}{9882} (\bibinfo{year}{2003}).

\bibitem[{\citenamefont{Steinhardt et~al.}(1983)\citenamefont{Steinhardt,
  Nelson, and Ronchetti}}]{steinhardt:1983}
\bibinfo{author}{\bibfnamefont{P.~J.} \bibnamefont{Steinhardt}},
  \bibinfo{author}{\bibfnamefont{D.~R.} \bibnamefont{Nelson}},
  \bibnamefont{and}
  \bibinfo{author}{\bibfnamefont{M.}~\bibnamefont{Ronchetti}},
  \bibinfo{journal}{Phys. Rev. B} \textbf{\bibinfo{volume}{28}},
  \bibinfo{pages}{784} (\bibinfo{year}{1983}).

\bibitem[{\citenamefont{Allen and Tildesley}(1987)}]{allen:1987}
\bibinfo{author}{\bibfnamefont{M.~P.} \bibnamefont{Allen}} \bibnamefont{and}
  \bibinfo{author}{\bibfnamefont{D.~J.} \bibnamefont{Tildesley}},
  \emph{\bibinfo{title}{Computer Simulation of Liquids}}
  (\bibinfo{publisher}{Oxford University Press, New York},
  \bibinfo{year}{1987}).

\end{thebibliography}

\vspace{0.05in}
\noindent {\bf Acknowledgments} \hspace{0.05in} The authors acknowledge primary financial support from the NSF MRSEC
DMR-1119826 (KZ). This work also benefited from
the facilities and staff of the Yale University Faculty of Arts and
Sciences High Performance Computing Center and the NSF (Grant
No. CNS-0821132) that in part funded acquisition of the computational
facilities.

\noindent {\bf Author contributions} \hspace{0.05in}
K. Z., J. S., M. D. S. and C. S. O. designed the work; K. Z. performed the simulation; K.Z. and C. S. O wrote the paper.

\noindent {\bf Additional information} \hspace{0.05in}
The authors declare no competing financial interests.  Supplementary information is available online.





\end{document}